\documentclass[letterpaper,12pt]{article}
\usepackage{array}
\usepackage{dcolumn}
\usepackage{float}
\usepackage[numbers]{natbib}
\usepackage{graphicx}
\usepackage{color}
\usepackage{amsmath}
\usepackage{amsfonts}
\usepackage{amsopn}
\usepackage{amsthm}
\usepackage{lscape}
\usepackage{multirow}

\theoremstyle{definition}

\theoremstyle{remark}

\DeclareMathOperator{\pr}{P}
\DeclareMathOperator{\epn}{E}
\DeclareMathOperator{\var}{Var}

\DeclareMathOperator{\argmin}{arg\;min}
\newcommand{\bmW}{\boldsymbol W}

\newcommand{\bmO}{\boldsymbol O}

\newcommand{\bmbeta}{\mbox{\boldmath${\beta}$}}

\renewcommand{\baselinestretch}{1.8}

\begin{document}
\begin{center}
{\LARGE A Unified Approach to Covariate Adjustment for Survival Endpoints in Randomized Clinical Trials}
\end{center}

\vspace{0.3cm}
\begin{center}
Zhiwei Zhang$^*$, Ya Wang and Dong Xi\\
Biostatistics Innovation Group, Gilead Sciences, Foster City, California, USA\\
$^*$zhiwei.zhang6@gilead.com
\end{center}

\vspace{0.3cm}
\centerline{\bf Abstract}
Covariate adjustment aims to improve the statistical efficiency of randomized trials by incorporating information from baseline covariates. Popular methods for covariate adjustment include analysis of covariance for continuous endpoints and standardized logistic regression for binary endpoints. For survival endpoints, while some covariate adjustment methods have been developed for specific effect measures, they are not commonly used in practice for various reasons, including high demands for theoretical and methodological sophistication as well as computational skills. This article describes an augmentation approach to covariate adjustment for survival endpoints that is relatively easy to understand and widely applicable to different effect measures. This approach involves augmenting a given treatment effect estimator in a way that preserves interpretation, consistency, and asymptotic normality. The optimal augmentation term, which minimizes asymptotic variance, can be estimated using various statistical and machine learning methods. Simulation results demonstrate that the augmentation approach can bring substantial gains in statistical efficiency. This approach has been implemented in an R package named \texttt{sleete}, which is described in detail and illustrated with real data.


\section{Introduction}\label{intro}

Randomized clinical trials are widely considered the gold standard for evaluating the safety and effectiveness of medical treatments. The statistical efficiency of clinical trials can be improved by incorporating information from baseline covariates (i.e., pre-treatment patient characteristics related to clinical outcomes). This usage of baseline covariates, commonly known as covariate adjustment, has received a great deal of attention from practitioners and regulators \citep{fda23}. There is a large and growing literature on statistical methods for covariate adjustment in clinical trials \citep[e.g.,][]{t08,lt08,z08,r08,m09a,m09b,r10,t12,z15,d19,w19,zm19,z20,y23a,y23b}. The increased awareness of covariate adjustment has led to increased use of covariate-adjusted methods by pharmaceutical statisticians. For example, it is increasingly common to analyze binary endpoints using standardized logistic regression \citep{m09b}. For continuous endpoints, the analysis of covariance method is presumably the most popular method for covariate adjustment.

This article is focused on right-censored survival endpoints, for which there seems to be no clear consensus on what is the preferred method for covariate adjustment. It may be tempting to incorporate covariates using a proportional hazards (PH) model; however, this practice changes the estimand from a marginal hazard ratio (HR) into a conditional HR and is susceptible to model mis-specification. The aforementioned literature on covariate adjustment contains several methods developed specifically for survival endpoints, which aim to improve the power of the log-rank test \citep{m09a,y23b} and the precision of estimating the HR for treatment in a PH model without baseline covariates \citep{lt08}, differences between survival curves \citep{z15}, differences in restricted mean survival time \citep[RMST;][]{d19}, and Mann--Whitney-type effect measures such as the win-lose probability difference \citep{z20}. Possible reasons for their less-than-common use in practice include high demands for theoretical and methodological sophistication as well as computational skills. Some of these methods are developed for specific effect measures and not easily adaptable to others. It would be helpful to practitioners to provide a simple and unified approach that is relatively easy to understand, widely applicable to different effect measures, and readily available in a user-friendly software package.

This article describes such a unified approach to covariate adjustment for survival endpoints together with a flexible and easy-to-use implementation. The approach we focus on is a general-purpose augmentation approach in which an initial treatment effect estimator, which typically does not make use of covariate data, is augmented with a term involving treatment assignment and an arbitrary function of covariates \citep{t08,t12,zm19,z20}. If the initial estimator is consistent and asymptotically linear, the augmented estimator is also consistent, asymptotically linear and therefore asymptotically normal under mild regularity conditions. The optimal augmentation, which minimizes the asymptotic variance of an augmented estimator, has a simple characterization. To take advantage of this optimality result, one may use a working regression model or a machine learning method to estimate the optimal augmentation and use the estimated augmentation to construct an augmented estimator. The resulting estimator remains consistent and asymptotically normal under fairly general conditions \citep{zm19,z20}. When the estimated augmentation involves machine learning, a sample splitting procedure can be used to maintain asymptotic normality and obtain a cross-validated variance estimate \citep{z11,c18,k20}. This approach is implemented in an R package named \texttt{sleete} (super learner for efficient estimation of treatment effects), which provides an option for sample splitting and many choices for estimating the optimal augmentation.

As noted earlier, there are many possible effect measures for survival endpoints. The most commonly used one in current practice is probably the HR for treatment in a PH model without baseline covariates. The PH assumption is a strong assumption and should not be taken for granted. Concerns about non-PH have motivated the consideration of alternative effect measures such as differences in survival probability and RMST as well as Mann--Whitney-type effect measures. The augmentation approach described in this article and implemented in the \texttt{sleete} package is flexible enough to accommodate all of these effect measures as well as other, user-defined ones. In other words, we leave the choice of effect measure to the user and strive to improve efficiency in estimating whatever effect measure the user chooses to estimate. When we work with the HR, we regard the PH model as a working model and conceptualize the true estimand as the limit of the maximum partial likelihood estimator \citep{lw89}. The \texttt{sleete} package allows the user to define a novel effect measure implicitly by supplying an estimator that is known to be consistent and asymptotically linear.

The rest of the article is organized as follows. The augmentation approach is described in Section \ref{meth} and evaluated in a simulation study in Section \ref{sim}. The \texttt{sleete} package is described in Appendix A and illustrated with real data in Section \ref{ex}. The article ends with a discussion in Section \ref{disc}.

\section{Methodology}\label{meth}

Consider a randomized clinical trial comparing an experimental treatment ($A=1$) with a control treatment ($A=0$) with respect to the time $T$ to a certain failure event, which may be death or a composite of death with other events. Let $\bmW$ be a vector of baseline covariates that may be related to $T$ in one or both treatment groups. Randomization implies that $A$ is independent of $\bmW$. Due to limited follow-up and possible drop-out, the failure time $T$ may not be fully observed and may be censored by a censoring time $C$. The observable outcome data consist of $X=T\wedge C$ and $\Delta=I(T\le C)$, where $\wedge$ denotes minimum and $I(\cdot)$ is the indicator function. The observed data from an individual subject will be written as $\bmO=(\bmW,A,X,\Delta)$, whose joint distribution is constrained by the known independence between $A$ and $\bmW$. The observed trial data consist of $n$ independent copies of $\bmO$ denoted by $\bmO_i=(\bmW_i,A_i,X_i,\Delta_i)$, $i=1,\dots,n$.

Let $\theta$ be a chosen measure of the effect of $A=1$ versus $A=0$ on $T$, and let $\overline\theta$ be an initial estimator of $\theta$. The choice of $\theta$ is an important question to consider but is not the focus in this article. Our only suggestion on $\theta$ is to apply a log transformation when $\theta$ is positive-valued by definition (e.g., HR to log-HR). The initial estimator $\overline\theta$ is typically unadjusted for covariates, but may involve some adjustment for covariates for other reasons than efficiency improvement. For example, $\overline\theta$ may be a log-HR estimator based on a stratified PH model if stratification is considered important for interpretation, or an inverse probability weighted estimator that accounts for informative censoring \citep[e.g.,][]{d19}. We assume that $\overline\theta$ is consistent for $\theta$ and asymptotically linear in the sense that
$$
\sqrt n(\overline\theta-\theta)=\frac{1}{\sqrt n}\sum_{i=1}^n\psi(\bmO_i)+o_p(1),
$$
where $\psi(\bmO)$ is the influence function of $\overline\theta$. (If $\overline\theta$ does not involve covariate data, then $\psi(\bmO)$ does not depend on $\bmW$.) This assumption is satisfied by most treatment effect estimators that are commonly used in practice \citep[e.g.,][]{lw89,fh91}. The assumed asymptotic linearity implies that $\sqrt n(\overline\theta-\theta)$ converges to a normal distribution with mean 0 and variance $\var\{\psi(\bmO)\}$. Our objective is to improve efficiency over $\overline\theta$ by incorporating information from $\bmW$.

The information in $\bmW$ can be incorporated using an augmentation approach based on semiparametric theory \citep{t06}. Following \citet{t08} and \citet{t12}, an augmented estimator of $\theta$ may be obtained as
\begin{equation}\label{aug.est}
\widehat\theta(b)=\overline\theta-\frac1n\sum_{i=1}^n(A_i-\pi)b(\bmW_i),
\end{equation}
where $\pi=\pr(A=1)$ (known by design) and $b(\bmW)$ is an arbitrary function of $\bmW$ such that $\epn\{b(\bmW)^2\}<\infty$. For any fixed function $b$, we have $\epn\{(A-\pi)b(\bmW)\}=\epn(A-\pi)\epn\{b(\bmW)\}=0$ due to randomization; therefore, the augmentation term $n^{-1}\sum_{i=1}^n(A_i-\pi)b(\bmW_i)$ converges to 0 by the law of large numbers, and $\widehat\theta(b)$ remains consistent for $\theta$. It is also easy to see that $\widehat\theta(b)$ is asymptotically linear with influence function $\psi(\bmO)-(A-\pi)b(\bmW)$ and, therefore, asymptotically normal with asymptotic variance
\begin{equation}\label{asymp.var}
\sigma^2(b):=\var\{\psi(\bmO)-(A-\pi)b(\bmW)\}=\epn\left[\{\psi(\bmO)-(A-\pi)b(\bmW)\}^2\right].
\end{equation}
Within this class of augmented estimators, it makes sense to search for the most efficient estimator with the smallest asymptotic variance. It can be shown \citep{t08,z08} that $\sigma^2(b)$ is minimized by setting $b(\bmW)$ equal to
\begin{equation}\label{opt.aug}
b_{\text{opt}}(\bmW)=\epn\{\psi(\bmO)|A=1,\bmW\}
-\epn\{\psi(\bmO)|A=0,\bmW\}.
\end{equation}
The corresponding estimator, $\widehat\theta(b_{\text{opt}})$, is the most efficient among all estimators of the form \eqref{aug.est}. In a nonparametric model for $\bmO$ with $\pr(A=1|\bmW)=\pr(A=1)=\pi$ as the only constraint, $\widehat\theta(b_{\text{opt}})$ is the most efficient among all regular, asymptotically linear estimators of $\theta$ based on $\{\bmO_i,i=1,\dots,n\}$ \citep{t06}.

The optimal augmentation function $b_{\text{opt}}$ is not fully known but can be estimated from trial data. This can be done indirectly by substituting in \eqref{opt.aug} an estimate of $\epn\{\psi(\bmO)|A,\bmW\}$ \citep{t08,z08}. Another approach, which is more direct, is to estimate $b_{\text{opt}}$ as the minimizer of \eqref{asymp.var} \citep{r08,zm19}. The \texttt{sleete} package implements the latter approach, which is more closely aligned with the objective of estimating $\theta$ efficiently and has been found to outperform the first (indirect) approach frequently \citep{zm19}. The idea is to treat \eqref{asymp.var} as a risk function and choose $b$ to minimize an empirical version of \eqref{asymp.var}, say
\begin{equation}\label{emp.risk}
\frac1n\sum_{i=1}^n\left\{\widehat\psi(\bmO_i)
-(A_i-\pi)b(\bmW_i)\right\}^2,
\end{equation}
where $\widehat\psi$ is an estimate of $\psi$. The estimate $\widehat\psi$ may be obtained by substituting estimates of unknown parameters and functions into an analytical expression for $\psi$ \citep[e.g.,][Section 3]{t12}, or using a resampling-based empirical influence function approach \citep{eg83,o17}, which is a convenient alternative when the analytical approach is cumbersome to implement \citep{z20}. The \texttt{sleete} package implements the analytical approach for some commonly used effect measures (e.g., log-HR and differences in survival probability and RMST) and also provides an option for the empirical influence function approach, which can be used for user-defined effect measures. Note that \eqref{emp.risk} can be rewritten as a weighted sum of squares:
\begin{equation}\label{obj.fct}
\sum_{i=1}^n\frac{(A_i-\pi)^2}{n}\left\{
\frac{\widehat\psi(\bmO_i)}{A_i-\pi}-b(\bmW_i)\right\}^2,
\end{equation}
where $(A_i-\pi)^2/n$ is regarded as a weight and $\widehat\psi(\bmO_i)/(A_i-\pi)$ as a response variable. When $\pi=1/2$ (a common case in clinical trials), the weight is identical for all subjects. Thus, $b_{\text{opt}}$ can be estimated by minimizing a (weighted) sum of squares.

As a simple example, suppose $b$ is parameterized as a linear function: $b(\bmW;\bmbeta)=(1,\bmW')\bmbeta$, where $\bmbeta$ is a parameter vector for optimization. The objective function \eqref{obj.fct} can now be minimized with respect to $\bmbeta$ using a standard linear regression routine based on (weighted) least squares. Let $\widehat\bmbeta$ denote the minimizer and let $\widehat b(\bmW)=(1,\bmW')\widehat\bmbeta$. If $b_{\text{opt}}$ happens to take the same linear form, then $\widehat\bmbeta$ will converge to the true value of $\bmbeta$ and $\widehat b$ to $b_{\text{opt}}$. More generally, regardless of the true form of $b_{\text{opt}}$, $\widehat\bmbeta$ is expected to converge to
$$
\bmbeta^*=\argmin_{\bmbeta}\var\{\psi(\bmO)-(A-\pi)(1,\bmW')\bmbeta\}=\argmin_{\bmbeta}\sigma^2(b(\cdot;\bmbeta)).
$$
Accordingly, $\widehat b$ converges to $b^*=b(\cdot;\bmbeta^*)$, which minimizes $\sigma^2(b)$ among all linear functions of the specified form. The augmented estimator $\widehat\theta(\widehat b)$ is asymptotically equivalent to $\widehat\theta(b^*)$ and, therefore, asymptotically normal with the same asymptotic variance $\sigma^2(b^*)$. A consistent estimator of $\sigma^2(b^*)$ is given by 
$$
\widehat\sigma^2(\widehat b)=\frac1n\sum_{i=1}^n\left\{\widehat\psi(\bmO_i)
-(A_i-\pi)\widehat b(\bmW_i)\right\}^2.
$$

Beyond linear regression, it is possible to use nonparametric machine learning methods to estimate the optimal augmentation function. Indeed, any prediction algorithm that is able to minimize a weighted sum of squared prediction errors can be used to minimize \eqref{obj.fct} as an objective function. There are many such prediction algorithms available \citep{h09}, some of which may be more appropriate than others for a given application. Without knowing or assuming which algorithms perform best, we can combine multiple candidate algorithms in a data-driven manner using the super learning principle \citep{p11}. The resulting super learner is a weighted average of the candidate algorithms with weights optimized through cross-validation. The \texttt{sleete} package relies on the \texttt{SuperLearner} package for super learning and allows the user to choose candidate algorithms freely from a large collection of algorithms available in \texttt{SuperLearner}. It also allows the user to skip super learning and use just one pre-selected algorithm to estimate the optimal augmentation. It should be noted that the use of machine learning methods to estimate $b_{\text{opt}}$ does necessitate special considerations. Let $\widetilde b$ be a nonparametric estimator of $b_{\text{opt}}$ which converges to some limit function $b^{\star}$. The standard argument for demonstrating the asymptotic equivalence of $\widehat\theta(\widetilde b)$ and $\widehat\theta(b^{\star})$ \citep[e.g.,][Appendix]{zm19} requires that $\widetilde b$ belongs to a Donsker class \citep{vw96}, a condition that limits the complexity of $\widetilde b$. Additionally, the variance estimator $\widehat\sigma^2(\widetilde b)$ may suffer from a downward re-substitution bias \citep{t12,zm19}.

\begin{figure}[htbp]
\centering
\includegraphics[width=0.8\textwidth]{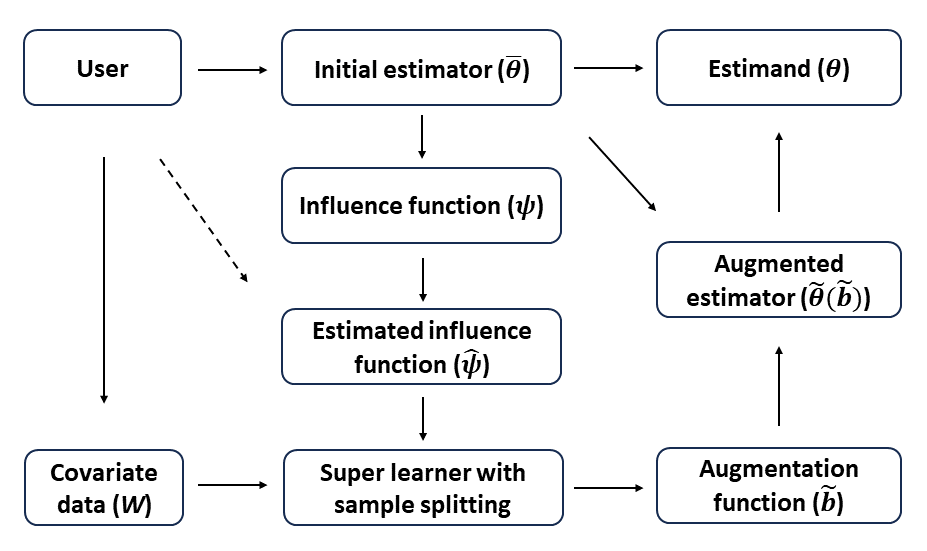}
\caption{A diagram for the augmentation approach based on the super learner together with sample splitting (which can be replaced by another learning method such as linear regression, with or without sample splitting). The dashed arrow indicates that the user can (but does not have to) provide an analytical estimate of the influence function for a user-defined treatment effect estimator.}
\label{diagram}
\end{figure}

Fortunately, both issues can be addressed easily using a simple sample splitting (also known as cross-fitting) procedure \citep{z11,c18,k20}. This procedure requires partitioning the study sample randomly into $K$ subsamples that are roughly equal in size. Let $\{V_i,i=1,\dots,n\}$ be a random sample from the uniform distribution on $\{1,\dots,K\}$. For each $k\in\{1,\dots,K\}$, let $\widehat\psi^{(-k)}$ and $\widetilde b^{(-k)}$ be obtained from $\{\bmO_i:V_i\not=k\}$ using the same methods for obtaining $\widehat\psi$ and $\widetilde b$. Then we can estimate $\theta$ with
$$
\widetilde\theta(\widetilde b)=\overline\theta-\frac1n\sum_{i=1}^n(A_i-\pi)\widetilde b^{(-V_i)}(\bmW_i),
$$
where $\widetilde b$ on the left-hand side denotes an algorithm (as opposed to an estimate). Without assuming a Donsker condition, it can be shown that $\widetilde\theta(\widetilde b)$ is consistent for $\theta$, asymptotically equivalent to $\widehat\theta(b^{\star})$, and therefore asymptotically normal with asymptotic variance $\sigma^2(b^{\star})$ \citep[Lemma 2]{k20}. In addition to removing the Donsker condition, the sample splitting procedure also produces a cross-validated estimator of the asymptotic variance $\sigma^2(b^{\star})$ of $\widetilde\theta(\widetilde b)$:
$$
\widetilde\sigma^2(\widetilde b)=\frac1{n}\sum_{i=1}^n\left\{\widehat\psi^{(-V_i)}(\bmO_i)
-(A_i-\pi)\widetilde b^{(-V_i)}(\bmW_i)\right\}^2.
$$
The cross-validated variance estimator is free of re-substitution bias and has been shown to provide adequate coverage \citep{zm19,z20}. The \texttt{sleete} package has an option for sample splitting, which is highly recommended when nonparametric machine learning methods are used to estimate $b_{\text{opt}}$. Figure \ref{diagram} provides a high-level visualization of this augmentation approach to covariate adjustment.

\section{Simulation}\label{sim}

This section reports a simulation study of the finite-sample performance of the methodology described in Section \ref{meth} in comparison with other methods. In this simulation study, the covariate vector $\bmW$ has three components $(W_1,W_2,W_3)$, which are independent and identically distributed as standard normal. The treatment indicator $A$ is a Bernoulli variable independent of $\bmW$ with $\pi=\pr(A=1)=1/2$ (unless otherwise stated). Given $(\bmW,A)$, the event time $T$ follows a Weibull distribution with shape parameter 3 and with conditional mean given below:
\begin{description}
\item{Scenario A: } $\epn(T|\bmW,A)=\Gamma(4/3)\exp(\gamma A+W_1+W_2-AW_2-AW_3)$;
\item{Scenario B: } $\epn(T|\bmW,A)=\Gamma(4/3)\exp(\gamma A+W_1+W_2-AW_2-AW_3+W_2W_3)$;
\item{Scenario C: } $\epn(T|\bmW,A)=\Gamma(4/3)\exp(1+\gamma A+W_1+W_2-AW_2-AW_3-W_1^2)$;
\item{Scenario D: } $\epn(T|\bmW,A)=\Gamma(4/3)\exp(1+\gamma A+W_1+W_2-AW_2-AW_3-W_1^2+W_2W_3)$,
\end{description}
where $\gamma=0$ or $1/2$ and $\Gamma(\cdot)$ is the Gamma function. The scale parameter in the above Weibull distribution is given by $\epn(T|\bmW,A)/\Gamma(4/3)$ in each scenario. It is easy to see that, when $\gamma=0$, the conditional distribution of $T$ given $A=1$ is identical to that given $A=0$ in each scenario. Thus, $\gamma=0$ is a null case with no marginal treatment effect, while $\gamma=1/2$ defines a non-null case where $(T|A=1)\sim\exp(1/2)(T|A=0)\approx1.65(T|A=0)$. Independently of $(\bmW,A,T)$, the censoring time $C$ follows a uniform distribution on the interval $(1,4)$, mimicking administrative censoring in a clinical trial with three years of patient accrual followed by another year of follow-up. The probability of observing an event, $\epn(\Delta)=\pr(T\le C)$, varies between 0.61 and 0.75 in the null and non-null cases of the four scenarios described above. Each simulation experiment is based on $10^4$ simulated trials.

We consider three marginal effect measures: the log-HR in a PH model for $(T|A)$, the difference in survival probability $\pr(T>\tau|A=1)-\pr(T>\tau|A=0)$, and the difference in RMST $\epn(T\wedge\tau|A=1)-\epn(T\wedge\tau|A=0)$, where $\tau=2$. When $\gamma=0$, the PH model for $(T|A)$ is correct, and the true value of $\theta$ is 0 in each scenario for each of the three effect measures considered here. When $\gamma=1/2$, the PH model is misspecified, and the true log-HR is defined as the large-sample limit of the maximum partial likelihood estimator (and approximated by analyzing a huge trial with $n=10^6$). Across the four scenarios, non-null values of $\theta$ range from $-0.36$ to $-0.34$ for the log-HR, from 0.12 to 0.13 for the difference in survival probability, and from 0.18 to 0.23 for the difference in RMST.

These effect measures will be estimated using unadjusted (for covariates) and augmented estimators with or without sample splitting. The unadjusted estimator of the log-HR is the maximum partial likelihood estimator. The unadjusted estimator of the difference in survival probablity is $\widehat S_1(\tau)-\widehat S_0(\tau)$, where $\widehat S_a(t)$ is the Kaplan-Meier estimator of $S_a(t)=\pr(T>t|A=a)$, $a=0,1$. The unadjusted estimator of the difference in RMST is
$$
\int_0^{\tau}\left\{\widehat S_1(t)-\widehat S_0(t)\right\}\/\text{d}t.
$$
These estimators may be augmented using a linear model given by $b(\bmW;\bmbeta)=(1,\bmW')\bmbeta$, an analogous additive model \citep{ht90}, a regression tree \citep{b84}, a random forest \citep{b01}, or a super learner that combines all of these methods. The super learner is based on five-fold cross-validation. The sample splitting, if conducted, is based on five-fold cross-fitting, which is external to the cross-validation in the super learner. For estimating the log-HR, two additional methods are included as comparators: the method of \citet{lt08} as implemented in the \texttt{speff2trial} package and the method of \citet{y23b} as implemented in the \texttt{RobinCar} package. These are abbreviated as LT08 and YSY23, respectively. Note that LT08 relies heavily on the PH assumption and may be inconsistent for our definition of log-HR under a misspecified PH model. On the other hand, YSY23 appears insensitive to model misspecification in the sense of converging to the same limit as the unadjusted estimator does.

The methods are compared in terms of empirical bias, standard deviation, relative efficiency, and coverage probability. The relative efficiency of an estimator is calculated as the inverse ratio of its variance to the variance of the unadjusted estimator. For example, an augmented estimator with relative efficiency 1.5 allows the sample size to be reduced by one third while maintaining the same level of precision attained by the unadjusted estimator.

We start with a moderate sample size ($n=250$) before exploring smaller ones. The non-null case (i.e., $\gamma=1/2$) simulation results for $n=250$ are reported in Table 1 (for the log-HR), Table 2 (for the difference in survival probability), and Table 3 (for the difference in RMST). In these tables, most estimators are virtually unbiased except for LT08 and the tree-based augmented estimator without sample splitting. The observed bias for LT08 is not surprising given that the method assumes PH. For the tree-based estimator, the observed bias is effectively reduced by sample splitting. The different estimators differ substantially in efficiency, and the unadjusted estimator is invariably the least efficient. For estimating the log-HR, the LT08, YSY23 and linearly augmented estimators are similarly efficient. The relative performance of the four individual augmentation methods (not including the super learner) seems to depend on the estimand and the scenario. For estimating the log-HR, for example, the linear model appears competitive in Scenario A, the additive model in Scenario C, and the random forest (with sample splitting) in Scenarios B--D. In most cases, the super learner with sample splitting performs similarly to, if not better than, the best-performing individual augmentation method. In terms of bias and variability, sample splitting appears to have little impact on the linearly and additively augmented estimators, reduce bias and increase variability for the tree-based estimator, and reduce variability without increasing bias for the random forest and super learner estimators. The results in Tables 1--3 also clearly demonstrate the benefit of sample splitting (or rather, the cross-validated variance estimator) in terms of variance estimation and coverage. Without sample splitting, the use of machine learning methods can result in poor coverage. With sample splitting, coverage is generally close to the nominal level.

The analogous null case (i.e., $\gamma=0$) simulation results are reported in Tables S1--3 in the Supplementary Materials. The results in Tables S1--3 are generally similar to those in Tables 1--3 except for smaller bias. For LT08, the bias reduction may be related to the fact that the PH model is correct when $\gamma=0$.

At smaller sample sizes (e.g., $n=200$), the use of machine learning methods produces erratic estimates in some samples ($<1$\%), making it difficult to compare methods in the same manner. The linear regression method appears more stable and continues to improve efficiency over the unadjusted estimator at sample sizes much lower than $n=250$. To illustrate this point, we report simulation results for $n=100$ in Table 4 (non-null case) and Table S4 (null case), where augmentation is based solely on linear regression, with or without sample splitting. The results in Tables 4 and S4 generally follow the same patterns observed previously for $n=250$. Note that LT08 and YSY23 can have sub-nominal coverage at $n=100$. For the linearly augmented estimator, sample splitting does not seem to improve efficiency, but it does clearly improve coverage. At $n=100$, even the unadjusted method can have under-coverage for estimating the difference in survival probability (and, to a lesser extent, the difference in RMST), and linear augmentation without sample splitting typically results in a small decrease in coverage. With sample splitting, the linearly augmented estimator provides near-nominal coverage for all three estimands in all situations considered.

All of the previous simulation results are based on $\pi=1/2$. As suggested by a reviewer, we have also conducted simulation experiments with $\pi=2/3$ for estimating the log-HR at $n=250$. The results, reported in Table S5 (null case) and  Table S6 (non-null case), are similar to those in Tables 1 and S1.

\section{Illustration}\label{ex}

\begin{figure}[htbp]
\centering
\includegraphics[width=0.65\textwidth]{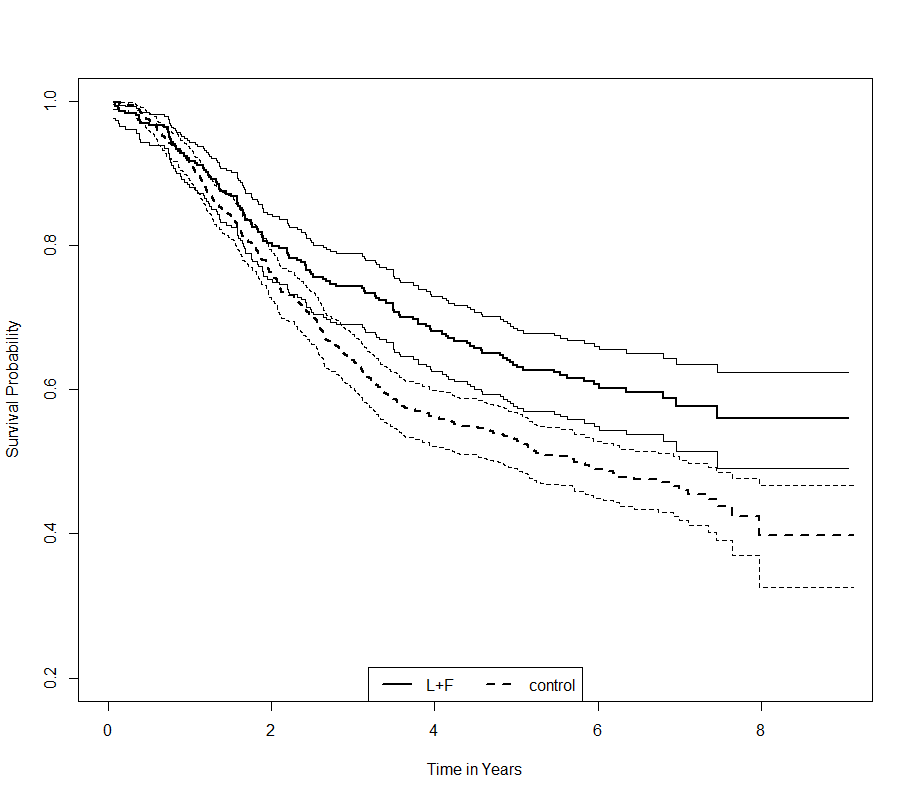}
\caption{Kaplan-Meier plots based on the colon cancer trial data: point estimates (thicker lines) and 95\% pointwise confidence intervals (thinner lines) for the L+F (solid lines) and control (dashed lines) groups.}
\label{KM.plots}
\end{figure}

We now illustrate the methods using a randomized clinical trial of adjuvant therapy regimens for preventing cancer recurrence and death after resection of stage III colon carcinoma \citep{m95}. The trial enrolled 929 eligible patients who had had curative-intent resections of stage III colon cancer in the previous one to five weeks, and randomly assigned them to observation only (control), levamisole alone or levamisole plus fluorouracil (L+F) with equal allocation among the three groups. The subjects were followed for a median duration of 6.5 years. We are interested in comparing the L+F regimen with control with respect to overall survival. Death was observed for 121 out of 304 patients in the L+F group and for 168 out of 315 patients in the control group. Figure \ref{KM.plots} shows Kaplan-Meier estimates of survival functions together with pointwise confidence intervals, reproduced with permission from \citet{z20}. Figure \ref{KM.plots} indicates that L+F has a delayed and long-lasting effect on overall survival, which raises questions about the appropriateness of the commonly used PH model. This is an important issue to consider in choosing an appropriate effect measure but is not a serious concern in our illustration.

The colon cancer trial data is readily available in the \texttt{survival} package. The available baseline covariates are sex, age, obstruction of colon by tumor (yes or no), perforation of colon (yes or no), adherence to nearby organs (yes or no), number of lymph nodes with detectable cancer, differentiation of tumor (well, moderate, or poor), extent of local spread (submucosa, muscle, serosa, or contiguous structures), and time from surgery to registration (long or short). The original dataset has duplicate entries for each subject to accommodate two event types: cancer recurrence and death. To create an appropriate dataset for our analysis, we will exclude duplicate entries for cancer recurrence as well as subjects in the levamisole-only group. 

\begin{verbatim}
> library(survival)
> data(colon)
> dim(colon); names(colon)
\end{verbatim}

\begin{verbatim}
[1] 1858   16
 [1] "id"       "study"    "rx"       "sex"      "age"      "obstruct"
 [7] "perfor"   "adhere"   "nodes"    "status"   "differ"   "extent" 
[13] "surg"     "node4"    "time"     "etype"   
\end{verbatim}

\begin{verbatim}
> data = subset(colon, subset=((etype==2)&(rx!="Lev")))
> dim(data)
\end{verbatim}

\begin{verbatim}
[1] 619  16
\end{verbatim}

The data will be analyzed using the \texttt{sleete} package, which is described in Appendix A. The \texttt{sleete} package provides two functions for data analysis: \texttt{mleete()} and \texttt{sleete()}. The former uses one specified learning algorithm to estimate the optimal augmentation, while the latter combines a specified collection of algorithms into a super learner. We first illustrate the \texttt{mleete()} function using three built-in methods for survival endpoints: \texttt{"log.HR"} (log-HR), \texttt{"surv.diff"} (difference in survival probability), and \texttt{"RMST.diff"} (difference in RMST). In the last two methods, the time argument \texttt{tau} is set to five years, or rather, $5\times365$ days. Otherwise we adopt the default settings of the \texttt{mleete()} function, including the use of linear regression with sample splitting.

\begin{verbatim}
> data$trt = as.numeric(data$rx=="Lev+5FU")
> pi = 0.5; tau = 5*365
> set.seed(12345)
> round(mleete(data, "time", event="status", trt="trt",
+              basecov.cont=c("age", "nodes", "differ", "extent"),
+              basecov.cat=c("sex", "obstruct", "perfor", "adhere", "surg", "node4"),
+              pi=pi, method="log.HR"), digits=3)
\end{verbatim}

\begin{verbatim}
           Pt. Est. Std. Err.
Unadjusted   -0.385     0.121
Augmented    -0.333     0.116
\end{verbatim}

\begin{verbatim}
> round(mleete(data, "time", event="status", trt="trt",
+              basecov.cont=c("age", "nodes", "differ", "extent"),
+              basecov.cat=c("sex", "obstruct", "perfor", "adhere", "surg", "node4"),
+              pi=pi, method="surv.diff", tau=tau), digits=3)
\end{verbatim}

\begin{verbatim}
           Pt. Est. Std. Err.
Unadjusted    0.116     0.040
Augmented     0.092     0.039
\end{verbatim}

\begin{verbatim}
> round(mleete(data, "time", event="status", trt="trt",
+              basecov.cont=c("age", "nodes", "differ", "extent"),
+              basecov.cat=c("sex", "obstruct", "perfor", "adhere", "surg", "node4"),
+              pi=pi, method="RMST.diff", tau=tau), digits=1)
\end{verbatim}

\begin{verbatim}
           Pt. Est. Std. Err.
Unadjusted      119.0      47.6
Augmented        97.3      44.9
\end{verbatim}

Next, we apply the \texttt{sleete()} function with the default settings, including the use of sample splitting and the same collection of candidate algorithms employed in the simulation study.

\begin{verbatim}
> round(sleete(data, "time", event="status", trt="trt",
+              basecov.cont=c("age", "nodes", "differ", "extent"),
+              basecov.cat=c("sex", "obstruct", "perfor", "adhere", "surg", "node4"),
+              pi=pi, method=log.HR), digits=3)
\end{verbatim}

\begin{verbatim}
                Pt. Est. Std. Err.
Unadjusted        -0.385     0.121
SL.lm             -0.308     0.117
SL.gam            -0.307     0.117
SL.rpart          -0.339     0.124
SL.randomForest   -0.307     0.119
SL                -0.306     0.118
\end{verbatim}

\begin{verbatim}
> round(sleete(data, "time", event="status", trt="trt",
+              basecov.cont=c("age", "nodes", "differ", "extent"),
+              basecov.cat=c("sex", "obstruct", "perfor", "adhere", "surg", "node4"),
+              pi=pi, method="surv.diff", tau=tau), digits=3)
\end{verbatim}

\begin{verbatim}
                Pt. Est. Std. Err.
Unadjusted         0.116     0.040
SL.lm              0.096     0.038
SL.gam             0.096     0.039
SL.rpart           0.099     0.040
SL.randomForest    0.103     0.039
SL                 0.096     0.039
\end{verbatim}

\begin{verbatim}
> round(sleete(data, "time", event="status", trt="trt",
+              basecov.cont=c("age", "nodes", "differ", "extent"),
+              basecov.cat=c("sex", "obstruct", "perfor", "adhere", "surg", "node4"),
+              pi=pi, method="RMST.diff", tau=tau), digits=1)
\end{verbatim}

\begin{verbatim}
                Pt. Est. Std. Err.
Unadjusted         119.0      47.6
SL.lm               88.0      45.0
SL.gam              88.3      45.0
SL.rpart            85.5      46.3
SL.randomForest     99.9      46.2
SL                  93.9      45.1
\end{verbatim}

\section{Discussion}\label{disc}

The augmentation methodology described in this article assumes randomized treatment assignment and the availability of an initial estimator that is consistent and asymptotically linear. The initial estimator may be designed to estimate some causal estimand under certain assumptions (e.g., PH). When the required assumptions fail to hold, the initial estimator may become inconsistent for the original causal estimand but typically converges to some quantity in an asymptotically linear fashion \citep[e.g.][]{lw89}. In this situation, the augmentation methodology can still be applied but the estimand needs to be redefined as the limit of the initial estimator. The interpretability of the redefined estimand will depend on the severity of assumption violations. For example, when the PH assumption is violated slightly, the limit of the maximum partial likelihood estimator may still be interpretable as an approximate log-HR. The choice of the initial estimator (and the implied estimand) is an important question which is largely separate from the efficient estimation problem considered here.

The simulation results presented here and in \citet{zm19} and \citet{z20} demonstrate that nonparametric machine learning methods can be quite useful for covariate adjustment, especially when the optimal augmentation function is highly non-linear. On the other hand, the use of such methods (as opposed to linear regression) does require some familiarity with machine learning algorithms and introduce some complications, including greater computational burden, choice of candidate algorithms, and choice of tuning parameters. Both the super learner and the sample splitting procedure introduce randomness into the results (point estimates and standard errors), which will require careful consideration in a regulatory setting. These complications may have hampered the use of machine learning methods for covariate adjustment. For instance, \citet{fda23} states specifically that the guidance does not address the use of machine learning methods for covariate adjustment. Clearly, much work remains to be done before nonparametric machine learning methods become widely used for covariate adjustment in the pharmaceutical industry.

Based on these practical considerations, we recommend the use of linear regression (as described in Section \ref{meth}) for covariate adjustment in registration trials. This method is similar in complexity to the analysis of covariance method for continous endpoints and the standardized logistic regression method for binary endpoints, and should not be considered a machine learning method in the sense of \citet{fda23}. The method is readily available in the \texttt{sleete} package and can be applied by calling the \texttt{mleete()} function with default settings. We hope that the additional availability of machine learning methods in the \texttt{sleete} package will facilitate further evaluation and research on these methods toward a better understanding of their performance and operating characteristics in realistic settings.

While the simulation study in Section \ref{sim} is limited to three commonly used effect measures for survival endpoints, the augmentation methodology in Section \ref{meth} and the \texttt{sleete} package can be applied to any (consistent and asymptotically linear) initial estimator and any (survival or non-survival) endpoint. As indicated in Appendix A, the \texttt{sleete} package currently has two other built-in methods for survival endpoints: \texttt{"log.HR.strat"} (log-HR in a stratified PH model) and \texttt{"MW.cens"} (Mann--Whitney-type effects for survival endpoints \citep{z20}). The package also has four built-in methods for non-survival endpoints: \texttt{"mean.diff"} (difference in mean or proportion), \texttt{"log.ratio"} (log-ratio of two means or proportions), \texttt{"log.OR"} (log-odds-ratio), and \texttt{"MW"} (Mann--Whitney-type effects without censoring \citep{z19}). Appendix A provides example code for method definition, which can be used to define a novel effect measure for an arbitrary endpoint.

\pagebreak
\renewcommand{\baselinestretch}{1.1}
\begin{table}[htbp]
{\small
\caption{Simulation results for $n=250$: empirical bias, standard deviation (SD), relative efficiency (RE) and coverage probability (CP) for estimating a {\bf non-null log-HR}.}\label{sim.250.log.HR}
\newcolumntype{d}{D{.}{.}{2}}
\newcolumntype{e}{D{.}{.}{3}}
\begin{center}
\begin{tabular}{lleeddceedd}
\hline
\hline
\multicolumn{1}{l}{Method}&\multicolumn{1}{l}{Augmentation}&\multicolumn{4}{c}{Without Sample Splitting}
&\multicolumn{1}{c}{}&\multicolumn{4}{c}{With Sample Splitting}\\
\cline{3-6}\cline{8-11}
&&\multicolumn{1}{c}{Bias}&\multicolumn{1}{c}{SD}&\multicolumn{1}{c}{RE}&\multicolumn{1}{c}{CP}
&\multicolumn{1}{c}{}&\multicolumn{1}{c}{Bias}&\multicolumn{1}{c}{SD}&\multicolumn{1}{c}{RE}&\multicolumn{1}{c}{CP}\\
\hline
\multicolumn{11}{c}{Scenario A}\\
unadjusted&&-0.001&0.156&1.00&0.95&&&&&\\
LT08&&0.006&0.108&2.09&0.93&&&&&\\
YSY23&&-0.002&0.105&2.22&0.95&&&&&\\
augmented&linear model&-0.002&0.105&2.21&0.95&&-0.002&0.105&2.19&0.96\\
augmented&additive model&-0.003&0.104&2.22&0.94&&-0.002&0.105&2.20&0.96\\
augmented&regression tree&0.012&0.118&1.75&0.87&&-0.002&0.124&1.58&0.95\\
augmented&random forest&-0.001&0.124&1.57&0.58&&-0.002&0.110&2.00&0.95\\
augmented&super learner&-0.001&0.108&2.09&0.86&&-0.002&0.105&2.21&0.95\\
\hline
\multicolumn{11}{c}{Scenario B}\\
unadjusted&&0.000&0.155&1.00&0.95&&&&&\\
LT08&&0.005&0.121&1.65&0.94&&&&&\\
YSY23&&-0.002&0.120&1.66&0.94&&&&&\\
augmented&linear model&-0.001&0.120&1.67&0.94&&-0.001&0.121&1.65&0.95\\
augmented&additive model&-0.003&0.120&1.68&0.94&&-0.001&0.121&1.65&0.95\\
augmented&regression tree&0.014&0.118&1.74&0.88&&-0.001&0.126&1.51&0.95\\
augmented&random forest&0.001&0.123&1.58&0.59&&-0.001&0.110&2.00&0.95\\
augmented&super learner&0.001&0.121&1.65&0.66&&-0.001&0.110&2.00&0.95\\
\hline
\multicolumn{11}{c}{Scenario C}\\
unadjusted&&-0.003&0.165&1.00&0.95&&&&&\\
LT08&&0.008&0.143&1.34&0.94&&&&&\\
YSY23&&-0.004&0.142&1.36&0.94&&&&&\\
augmented&linear model&-0.004&0.141&1.36&0.95&&-0.003&0.142&1.35&0.95\\
augmented&additive model&-0.005&0.122&1.84&0.94&&-0.003&0.121&1.85&0.96\\
augmented&regression tree&0.016&0.129&1.65&0.86&&-0.003&0.133&1.54&0.95\\
augmented&random forest&-0.001&0.133&1.54&0.59&&-0.003&0.120&1.90&0.95\\
augmented&super learner&-0.001&0.124&1.77&0.80&&-0.003&0.118&1.97&0.95\\
\hline
\multicolumn{11}{c}{Scenario D}\\
unadjusted&&-0.001&0.165&1.00&0.95&&&&&\\
LT08&&0.009&0.148&1.25&0.94&&&&&\\
YSY23&&-0.002&0.148&1.24&0.94&&&&&\\
augmented&linear model&-0.002&0.148&1.25&0.95&&-0.001&0.149&1.24&0.95\\
augmented&additive model&-0.003&0.131&1.60&0.94&&0.000&0.131&1.60&0.96\\
augmented&regression tree&0.020&0.122&1.84&0.88&&-0.003&0.128&1.66&0.96\\
augmented&random forest&0.002&0.130&1.62&0.58&&-0.002&0.115&2.08&0.96\\
augmented&super learner&0.003&0.128&1.68&0.65&&-0.002&0.115&2.07&0.96\\
\hline
\end{tabular}
\end{center}
}
\end{table}

\pagebreak
\renewcommand{\baselinestretch}{1.1}
\begin{table}[htbp]
{\small
\caption{Simulation results for $n=250$: empirical bias, standard deviation (SD), relative efficiency (RE) and coverage probability (CP) for estimating a {\bf non-null difference in survival probability}.}\label{sim.250.surv.diff}
\newcolumntype{d}{D{.}{.}{2}}
\newcolumntype{e}{D{.}{.}{3}}
\begin{center}
\begin{tabular}{lleeddceedd}
\hline
\hline
\multicolumn{1}{l}{Method}&\multicolumn{1}{l}{Augmentation}&\multicolumn{4}{c}{Without Sample Splitting}
&\multicolumn{1}{c}{}&\multicolumn{4}{c}{With Sample Splitting}\\
\cline{3-6}\cline{8-11}
&&\multicolumn{1}{c}{Bias}&\multicolumn{1}{c}{SD}&\multicolumn{1}{c}{RE}&\multicolumn{1}{c}{CP}
&\multicolumn{1}{c}{}&\multicolumn{1}{c}{Bias}&\multicolumn{1}{c}{SD}&\multicolumn{1}{c}{RE}&\multicolumn{1}{c}{CP}\\
\hline
\multicolumn{11}{c}{Scenario A}\\
unadjusted&&0.000&0.064&1.00&0.94&&&&&\\
augmented&linear model&0.001&0.051&1.58&0.94&&0.001&0.051&1.58&0.95\\
augmented&additive model&0.001&0.050&1.61&0.93&&0.001&0.050&1.62&0.95\\
augmented&regression tree&-0.014&0.050&1.61&0.87&&0.000&0.055&1.31&0.94\\
augmented&random forest&0.000&0.054&1.38&0.61&&0.000&0.051&1.56&0.94\\
augmented&super learner&-0.001&0.050&1.59&0.83&&0.000&0.050&1.63&0.95\\
\hline
\multicolumn{11}{c}{Scenario B}\\
unadjusted&&0.000&0.063&1.00&0.94&&&&&\\
augmented&linear model&0.000&0.054&1.35&0.94&&0.000&0.054&1.34&0.95\\
augmented&additive model&0.001&0.054&1.37&0.94&&0.000&0.054&1.37&0.95\\
augmented&regression tree&-0.014&0.050&1.58&0.87&&0.000&0.056&1.27&0.95\\
augmented&random forest&-0.001&0.054&1.37&0.61&&0.000&0.051&1.53&0.95\\
augmented&super learner&-0.001&0.053&1.44&0.73&&0.000&0.051&1.54&0.95\\
\hline
\multicolumn{11}{c}{Scenario C}\\
unadjusted&&0.001&0.066&1.00&0.94&&&&&\\
augmented&linear model&0.001&0.059&1.24&0.94&&0.001&0.059&1.24&0.95\\
augmented&additive model&0.002&0.055&1.45&0.94&&0.001&0.054&1.46&0.95\\
augmented&regression tree&-0.014&0.053&1.52&0.87&&0.001&0.058&1.27&0.94\\
augmented&random forest&0.000&0.056&1.36&0.62&&0.001&0.054&1.50&0.94\\
augmented&super learner&-0.001&0.054&1.47&0.80&&0.001&0.053&1.54&0.95\\
\hline
\multicolumn{11}{c}{Scenario D}\\
unadjusted&&0.000&0.066&1.00&0.94&&&&&\\
augmented&linear model&0.001&0.061&1.15&0.94&&0.000&0.061&1.15&0.95\\
augmented&additive model&0.001&0.057&1.33&0.93&&0.000&0.057&1.33&0.95\\
augmented&regression tree&-0.014&0.051&1.65&0.88&&0.000&0.057&1.34&0.94\\
augmented&random forest&-0.001&0.055&1.42&0.61&&0.000&0.052&1.61&0.94\\
augmented&super learner&-0.002&0.054&1.47&0.69&&0.000&0.052&1.60&0.95\\
\hline
\end{tabular}
\end{center}
}
\end{table}

\pagebreak
\renewcommand{\baselinestretch}{1.1}
\begin{table}[htbp]
{\small
\caption{Simulation results for $n=250$: empirical bias, standard deviation (SD), relative efficiency (RE) and coverage probability (CP) for estimating a {\bf non-null difference in RMST}.}\label{sim.250.rmst.diff}
\newcolumntype{d}{D{.}{.}{2}}
\newcolumntype{e}{D{.}{.}{3}}
\begin{center}
\begin{tabular}{lleeddceedd}
\hline
\hline
\multicolumn{1}{l}{Method}&\multicolumn{1}{l}{Augmentation}&\multicolumn{4}{c}{Without Sample Splitting}
&\multicolumn{1}{c}{}&\multicolumn{4}{c}{With Sample Splitting}\\
\cline{3-6}\cline{8-11}
&&\multicolumn{1}{c}{Bias}&\multicolumn{1}{c}{SD}&\multicolumn{1}{c}{RE}&\multicolumn{1}{c}{CP}
&\multicolumn{1}{c}{}&\multicolumn{1}{c}{Bias}&\multicolumn{1}{c}{SD}&\multicolumn{1}{c}{RE}&\multicolumn{1}{c}{CP}\\
\hline
\multicolumn{11}{c}{Scenario A}\\
unadjusted&&0.000&0.096&1.00&0.94&&&&&\\
augmented&linear model&0.002&0.063&2.30&0.94&&0.001&0.063&2.30&0.95\\
augmented&additive model&0.002&0.063&2.31&0.93&&0.001&0.063&2.31&0.95\\
augmented&regression tree&-0.008&0.072&1.78&0.85&&0.000&0.075&1.65&0.94\\
augmented&random forest&0.000&0.076&1.60&0.56&&0.001&0.066&2.11&0.94\\
augmented&super learner&0.001&0.066&2.13&0.83&&0.001&0.063&2.33&0.94\\
\hline
\multicolumn{11}{c}{Scenario B}\\
unadjusted&&-0.001&0.097&1.00&0.94&&&&&\\
augmented&linear model&0.000&0.072&1.83&0.94&&-0.001&0.072&1.81&0.95\\
augmented&additive model&0.001&0.072&1.84&0.93&&-0.001&0.072&1.82&0.94\\
augmented&regression tree&-0.011&0.070&1.91&0.87&&0.000&0.075&1.68&0.95\\
augmented&random forest&-0.002&0.075&1.67&0.56&&0.000&0.064&2.30&0.94\\
augmented&super learner&-0.002&0.074&1.75&0.62&&0.000&0.064&2.30&0.94\\
\hline
\multicolumn{11}{c}{Scenario C}\\
unadjusted&&0.000&0.099&1.00&0.94&&&&&\\
augmented&linear model&0.001&0.084&1.40&0.94&&0.001&0.084&1.38&0.95\\
augmented&additive model&0.001&0.067&2.20&0.93&&0.000&0.067&2.22&0.95\\
augmented&regression tree&-0.013&0.067&2.19&0.86&&0.001&0.071&1.95&0.94\\
augmented&random forest&-0.001&0.076&1.72&0.56&&0.000&0.064&2.38&0.94\\
augmented&super learner&-0.001&0.069&2.07&0.76&&0.000&0.063&2.46&0.94\\
\hline
\multicolumn{11}{c}{Scenario D}\\
unadjusted&&-0.001&0.102&1.00&0.95&&&&&\\
augmented&linear model&0.000&0.088&1.34&0.94&&0.000&0.088&1.32&0.95\\
augmented&additive model&0.000&0.072&1.98&0.93&&-0.001&0.073&1.95&0.95\\
augmented&regression tree&-0.017&0.064&2.54&0.87&&0.000&0.070&2.12&0.95\\
augmented&random forest&-0.002&0.075&1.85&0.54&&0.000&0.061&2.76&0.95\\
augmented&super learner&-0.003&0.073&1.95&0.60&&0.000&0.061&2.75&0.95\\
\hline
\end{tabular}
\end{center}
}
\end{table}

\renewcommand{\baselinestretch}{1.1}
\begin{table}[htbp]
{\small
\caption{Simulation results for $n=100$: empirical bias, standard deviation (SD), relative efficiency (RE) and coverage probability (CP) for all three non-null estimands.}\label{sim.100}
\newcolumntype{d}{D{.}{.}{2}}
\newcolumntype{e}{D{.}{.}{3}}
\begin{center}
\begin{tabular}{lleeddceedd}
\hline
\hline
\multicolumn{1}{l}{Scenario}&\multicolumn{1}{l}{Method}&\multicolumn{4}{c}{Without Sample Splitting}
&\multicolumn{1}{c}{}&\multicolumn{4}{c}{With Sample Splitting}\\
\cline{3-6}\cline{8-11}
&&\multicolumn{1}{c}{Bias}&\multicolumn{1}{c}{SD}&\multicolumn{1}{c}{RE}&\multicolumn{1}{c}{CP}
&\multicolumn{1}{c}{}&\multicolumn{1}{c}{Bias}&\multicolumn{1}{c}{SD}&\multicolumn{1}{c}{RE}&\multicolumn{1}{c}{CP}\\
\hline
\multicolumn{11}{c}{$\theta=\text{log-HR}$}\\
A&unadjusted&-0.006&0.250&1.00&0.95&&&&&\\
&LT08&0.012&0.177&1.98&0.92&&&&&\\
&YSY23&-0.007&0.168&2.20&0.93&&&&&\\
&augmented&-0.005&0.170&2.17&0.94&&-0.005&0.171&2.13&0.96\\
B&unadjusted&-0.009&0.249&1.00&0.95&&&&&\\
&LT08&0.002&0.194&1.66&0.93&&&&&\\
&YSY23&-0.012&0.191&1.70&0.94&&&&&\\
&augmented&-0.011&0.191&1.70&0.95&&-0.010&0.194&1.65&0.96\\
C&unadjusted&-0.005&0.268&1.00&0.94&&&&&\\
&LT08&0.016&0.234&1.31&0.92&&&&&\\
&YSY23&-0.008&0.233&1.32&0.93&&&&&\\
&augmented&-0.007&0.231&1.34&0.93&&-0.004&0.233&1.32&0.96\\
D&unadjusted&-0.005&0.267&1.00&0.95&&&&&\\
&LT08&0.015&0.238&1.26&0.93&&&&&\\
&YSY23&-0.009&0.241&1.22&0.93&&&&&\\
&augmented&-0.008&0.239&1.25&0.94&&-0.005&0.242&1.22&0.96\\
\hline
\multicolumn{11}{c}{$\theta=\text{difference in survival probability}$}\\
A&unadjusted&0.001&0.101&1.00&0.93&&&&&\\
&augmented&0.001&0.082&1.54&0.91&&0.000&0.082&1.54&0.94\\
B&unadjusted&0.001&0.101&1.00&0.93&&&&&\\
&augmented&0.002&0.087&1.34&0.92&&0.001&0.087&1.33&0.94\\
C&unadjusted&0.000&0.106&1.00&0.93&&&&&\\
&augmented&0.001&0.097&1.20&0.91&&0.000&0.097&1.20&0.94\\
D&unadjusted&-0.001&0.105&1.00&0.93&&&&&\\
&augmented&0.000&0.098&1.14&0.92&&-0.001&0.098&1.14&0.94\\
\hline
\multicolumn{11}{c}{$\theta=\text{difference in RMST}$}\\
A&unadjusted&0.002&0.153&1.00&0.93&&&&&\\
&augmented&0.003&0.103&2.21&0.92&&0.001&0.103&2.22&0.95\\
B&unadjusted&0.002&0.154&1.00&0.94&&&&&\\
&augmented&0.004&0.115&1.81&0.92&&0.002&0.116&1.78&0.94\\
C&unadjusted&0.000&0.160&1.00&0.93&&&&&\\
&augmented&0.001&0.136&1.38&0.92&&0.000&0.138&1.35&0.94\\
D&unadjusted&0.001&0.161&1.00&0.93&&&&&\\
&augmented&0.001&0.139&1.36&0.93&&0.000&0.141&1.31&0.94\\
\hline
\end{tabular}
\end{center}
}
\end{table}

\end{document}